# Highly Accelerated Multishot EPI through Synergistic Machine Learning and Joint Reconstruction

Berkin Bilgic[1,2,3], Itthi Chatnuntawech[4], Mary Kate Manhard[1,2], Qiyuan Tian[1,2], Congyu Liao[1,2], Stephen F. Cauley[1,2], Susie Y. Huang[1,2,3], Jonathan R. Polimeni[1,2,3], Lawrence L. Wald[1,2,3], Kawin Setsompop[1,2,3]

1 Athinoula A. Martinos Center for Biomedical Imaging, Charlestown, MA, USA
2 Department of Radiology, Harvard Medical School, Boston, MA, USA
3 Harvard-MIT Health Sciences and Technology, MIT, Cambridge, MA, USA
4 National Nanotechnology Center, National Science and Technology Development Agency, Pathum Thani, Thailand

**Corresponding author:**
Itthi Chatnuntawech
itthi.cha@nanotec.or.th

# ABSTRACT

**Purpose:** To introduce a combined machine learning (ML) and physics-based image reconstruction framework that enables navigator-free, highly accelerated multishot echo planar imaging (msEPI), and demonstrate its application in high-resolution structural and diffusion imaging.

**Methods:** Singleshot EPI is an efficient encoding technique, but does not lend itself well to high-resolution imaging due to severe distortion artifacts and blurring. While msEPI can mitigate these artifacts, high-quality msEPI has been elusive because of phase mismatch arising from shot-to-shot variations which preclude the combination of the multiple-shot data into a single image. We employ deep learning to obtain an interim image with minimal artifacts, which permits estimation of image phase variations due to shot-to-shot changes. These variations are then included in a Joint Virtual Coil Sensitivity Encoding (JVC-SENSE) reconstruction to utilize data from all shots and improve upon the ML solution.

**Results:** Our combined ML + physics approach enabled $R_{inplane}$ x MultiBand (MB) = 8x2-fold acceleration using 2 EPI-shots for multi-echo imaging, so that whole-brain $T_2$ and $T_2$* parameter maps could be derived from an 8.3 sec acquisition at 1x1x3mm$^3$ resolution. This has also allowed high-resolution diffusion imaging with high geometric fidelity using 5-shots at $R_{inplane}$ x MB = 9x2-fold acceleration. To make these possible, we extended the state-of-the-art MUSSELS reconstruction technique to Simultaneous MultiSlice (SMS) encoding and used it as an input to our ML network.

**Conclusion:** Combination of ML and JVC-SENSE enabled navigator-free msEPI at higher accelerations than previously possible while using fewer shots, with reduced vulnerability to poor generalizability and poor acceptance of end-to-end ML approaches.

# INTRODUCTION

Slow image encoding has constrained clinical MRI scans to use 2-dimensional encoding and thick slices, often with slice gaps, so that whole-brain exams can be completed within acceptable time frames. In addition to the information loss, such inefficient acquisition poses a barrier to MRI evaluation of hospitalized patients who are critically ill and can neither hold still nor tolerate long scans. Low patient throughput due to inefficient imaging also increases the time from symptom onset to diagnosis, thereby delaying treatment.

To overcome the slow image encoding barrier, recent screening protocols have moved to singleshot Echo Planar Imaging (ssEPI) to provide multi-contrast information (1,2). Unfortunately, the reduced geometric fidelity of these protocols may confound/obscure localization of salient imaging findings. The problem arises from severe distortion and blurring artifacts in ssEPI at high in-plane resolutions, where a large area of k-space has to be covered within a single readout in the presence of $B_0$ inhomogeneity and $T_2^*$ signal decay. These effects are only partially mitigated at relatively high in-plane acceleration (e.g. $R_{inplane}$=3).

While multishot EPI (msEPI) can mitigate blurring and distortion, high-quality msEPI has been elusive because combining the multiple-shot data into a single image is prohibitively difficult, especially at high in-plane acceleration. Image phase mismatches between the shots caused by physiological variations (respiration, cardiac pulsation) or motion under the influence of diffusion encoding gradients lead to severe ghosting artifacts. To date, the application of msEPI has been restricted to diffusion imaging, where two types of solutions have been proposed to combine the shots: (i) navigator-based approaches that require additional data acquisition to capture shot-to-shot phase variations (3–7), and (ii) navigator-free techniques that estimate these variations from the data itself (8–11). In (ii), multiplexed sensitivity encoding (MUSE) (9) and its extensions (12,13), rely on parallel imaging to reconstruct an intermediate image for each shot independently to estimate the physiological variations before jointly reconstructing all multishot data together. This limits the achievable distortion and blurring reduction to 4 to 6-fold, since parallel imaging with modern RF receive coil arrays breaks down beyond such acceleration in the phase-encoding direction. MUSSELS, on the other hand, does not explicitly estimate the phase of each shot image, but employs sensitivity encoding and similarities across multishot data in the form of structured low-rank matrix completion (11,14). This has allowed MUSSELS to undersample the k-space of each shot by $R_{inplane}$=8-fold to reduce distortion and blurring artifacts as well as the echo time (TE). Images could be successfully reconstructed using 4-shots of data, so that the net acceleration factor became $R_{net}$ = 8/4 = 2-fold. It is important to note another class of navigator-free multishot diffusion imaging techniques, which utilize non-Cartesian trajectories that allow for estimation of low-resolution image phase information from the densely sampled portion of each shot (15,16). Such self-navigation property may come at the cost of blurring/distortion in the resulting images.

In this contribution, we introduce a new reconstruction framework that utilizes a synergistic combination of machine learning (ML) and physics (or forward-model) based reconstruction, and demonstrate its application in structural and diffusion msEPI with high geometric fidelity. We term our combined ML + physics approach Network Estimated Artifacts for Tempered Reconstruction (NEATR), and incorporate Simultaneous MultiSlice (SMS) for extra efficiency. To this end, we have extended MUSSELS to SMS encoding, and utilized the readout extended FOV concept (17) to seamlessly integrate slice acceleration into this framework. We start from SMS-MUSSELS reconstruction of highly accelerated msEPI using a smaller number of acquisition shots, and pass the intermediate solution through our deep neural network to mitigate the reconstruction artifacts from SMS-MUSSELS. Using this interim image with minimal artifacts allows us to solve for the image phase of each shot using phase-regularized parallel imaging (18). Given the phase of each shot, we then perform a Joint Virtual Coil Sensitivity Encoding (JVC-SENSE) reconstruction where we utilize the k-space data from all shots as well as virtual coil concept (19–21) to solve for the combined magnitude image.

We demonstrate the application of SMS-NEATR in spin-and-gradient-echo (SAGE (22)) msEPI acquisition at $R_{inplane}$ x MultiBand (MB) = 8x2-fold acceleration using 2-shots. Compared to the newly developed SMS-MUSSELS reconstruction, which is also used as an input to our network, we demonstrate ~30% improvement in root-mean-squared error (RMSE) in high-resolution structural images. We observe larger gains in ghosting/aliasing artifact mitigation in the harder problem of diffusion imaging, where SMS-NEATR allows for $R_{inplane}$ x MB = 9x2-fold acceleration using 5-shots.

These are made possible by the deep learning step that enables phase estimations at such high acceleration factors. Importantly, the final use of a rigorous physics-based forward-model reconstruction limits the role of ML in the final reconstruction. Thus, SMS-NEATR allows us to tap into the potential of convolutional neural networks (CNN) to solve for important nuisance modulations and unknowns in the forward model without treating the reconstruction as an end-to-end process. The result is a better harnessed sensitivity encoding with full utilization of the scanner hardware. Our strategy paves the way to reaping the benefits of ML while constraining potential damage from utilizing it on data beyond its training experience, and without being exposed to the vulnerabilities of not knowing exactly what the reconstruction is doing. Our approach of using ML to estimate nuisance parameters that are hard to determine could allow physics-based reconstructions to work well in other applications, such as retrospective motion-correction without navigation or additional hardware.

We provide Matlab source code and data to reproduce our diffusion msEPI results here: https://bit.ly/2QgBg9U

Supporting Information figures can be accessed here: https://bit.ly/2unY2iJ

## METHODS

### Reconstruction Overview

The SMS-NEATR flowchart is presented in **Fig. 1**. We begin by performing an SMS-MUSSELS reconstruction on the highly-accelerated (e.g. $R_{inplane}$ x MB = 8x2) msEPI data to obtain an initial image estimate with mitigated artifacts from the nuisance phase between shots. The image is further improved using U-Net processing (23) which estimates a refined image with minimal artifacts. Starting from this reconstruction, we estimate the phase image corresponding to each shot using phase-regularized parallel imaging (24). Given the estimated shot-to-shot phase variations, we then perform a physics-based joint reconstruction (JVC-SENSE) to arrive at the final solution. JVC-SENSE incorporates slice acceleration and uses k-space data from all shots and their conjugate symmetric counterparts to solve for a common magnitude image. We detail the individual steps next.

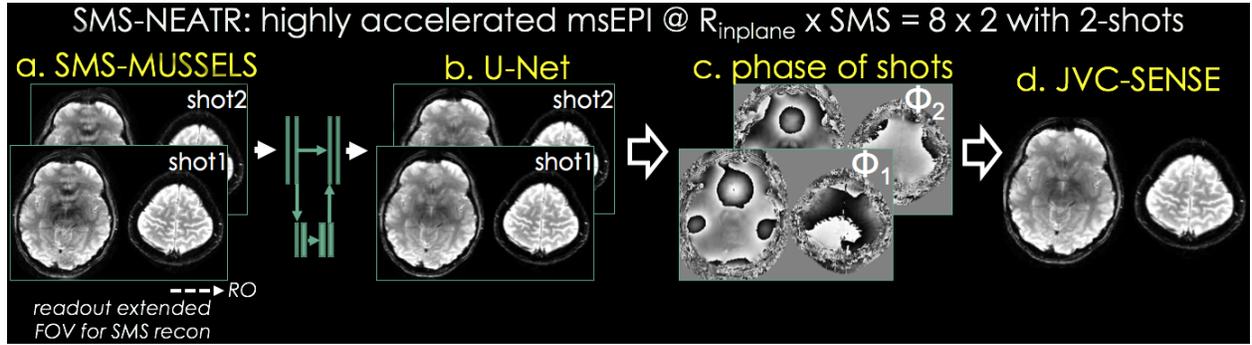

*Figure 1 SMS-NEATR is a combined machine learning and physics-based reconstruction technique for highly-accelerated msEPI acquisition. We developed SMS-MUSSELS algorithm to provide an initial solution, which may suffer from artifacts due to high acceleration ($R_{inplane}$xMB=8x2 with 2-shots). Starting from this, residual learning with U-Net architecture provides an interim image with minimal artifacts. Given this solution, phase cycling algorithm is used for estimating shot-phases, which are then utilized as sensitivity variations in a final joint virtual coil (JVC) SENSE reconstruction.*

### SMS-MUSSELS Formalism

The first step of SMS-NEATR is based on a MUSSELS reconstruction, where the input is the acquired multi-shot k-space data, and the output is an estimate of shot images which are further refined in the later steps. To begin with, we will ignore SMS encoding and consider only in-plane acceleration. In this case, MUSSELS entails the solution of the following optimization problem:

$$min_x \sum_{t=1}^{N_s} \|F_t C x_t - d_t\|_2^2 + \lambda \|\mathcal{H}(x)\|_* \quad \text{Eq1}$$

where $F_t$ represents the undersampled discrete Fourier transform (DFT) corresponding to shot $t$, $C$ are the coil sensitivities, $x_t$ is the unknown complex-valued image in shot $t$ with size $N_1 \times N_2$, and $d_t$ are the acquired k-space data in this shot. The term $\|F_t C x_t - d_t\|_2^2$ thus represents our data consistency through sensitivity encoding (25). The operator $\mathcal{H}(\cdot)$ first applies the DFT, and then extracts $r \times r \times N_s$ patches in k-space to generate a data matrix $\mathcal{H}(x)$ with block-wise Hankel structure (11,26–28). This operator acts on a 3-dimensional data structure $x$ of size $N_1 \times N_2 \times N_s$, which is formed by concatenating the images $x_t$ from all $N_s$ shots together. The nuclear norm constraint $\|\mathcal{H}(x)\|_*$ thus enforces a low-rank prior on the block-Hankel representation of the multishot data in k-space. This prior is similar to the SAKE formulation (28), albeit with two differences: the coil axis is now replaced by the shot dimension, and sensitivity encoding is explicitly exploited. As such, we follow the SAKE approach and pursue a simple, POCS-SENSE like algorithm (29) to solve Eq1 as detailed in the Appendix. We will show that the advanced FISTA update rules (30) improve convergence and image quality. We also note that the cost function being minimized differs from the convex optimization problem set by the original MUSSELS approach, and is more similar to LORAKS-type approaches (31) as they solve a non-convex problem.

Extension to SMS: We developed a new approach to allow MUSSELS to work with SMS encoding using the readout-extended FOV concept (17). This represents SMS as undersampling in the $k_x$ axis by concatenating the two slices along the *readout* (Fig.1a). In-plane and slice acceleration could thus be captured using the Fourier operator $F_t$ in Eq1, now with simultaneous $k_x$-$k_y$ undersampling.

In the next step of the SMS-NEATR reconstruction, we use the estimated shot images $\{x_t\}_{t=1}^{N_S}$ as input to a residual CNN (32,33) with U-Net architecture (23). The network aims to learn and mitigate the reconstruction errors in SMS-MUSSELS and provide output shot images, $\{u_t\}_{t=1}^{N_S}$, with minimal artifacts.

## Network Architecture

We used a patch-based U-Net to learn the mapping between the initial reconstruction and its difference to the ground truth image in a slice-by-slice manner. The network consisted of 5 levels **(Fig. 2)**, and the number of convolutional filters was 64 at the highest level. As the size of input was reduced 2-fold by max pooling in the next level, the number of filters was increased 2-fold to retain the total number of kernel weights in each level. The kernels had size 3x3, and each dropout layer set a randomly selected 5% of its input units to zero to help avoid overfitting (34). Leaky ReLU was selected as the nonlinear activation function (35). Batch normalization (BN) was utilized to help accelerate training and avoid saturating nonlinearities (36).

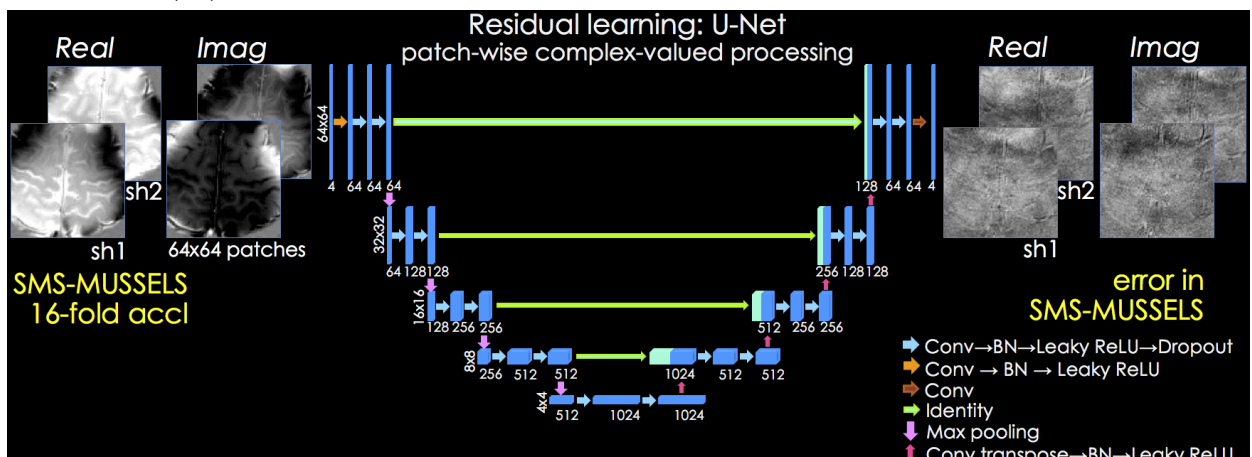

*Figure 2 U-Net architecture is used to learn the mapping between patches of shot-images reconstructed with SMS-MUSSELS, and their difference to reference data. Both the input and output have been decomposed into real and imaginary components to enable complex-valued processing for SAGE reconstruction. 64×64 patches from all the shots are presented as input to a 5-level network, where the first level uses 64 convolutional filters. To help provide scale invariance, max pooling operators downsample the patches after each layer. At the same time, the number of filters are doubled to retain the total number of kernel weights at each level.*

For SAGE reconstruction, we trained a complex-valued network by separating each of the 2-shot images into its real and imaginary components. This way we used 4 channels, $\{\Re(x_1), \Im(x_1), \Re(x_2), \Im(x_2)\}$, in this network configuration (Fig. 2). For DWI, we explored using both complex-valued and magnitude-based networks. The complex network made use of 10 channels (real and imaginary components from 5-shots) while the absolute value of each of the 5-shots, $\{|x_1|, \dots, |x_5|\}$ was used as input channels for the magnitude-based model. The remainder of the manuscript focuses on the magnitude-based DWI network, and the complex-valued U-Net results are reported in Supporting Information Fig S9 and S10.

### Network Implementation

Keras programming interface (37) with Tensorflow (38) backend were chosen to perform the training. ADAM optimizer (39) was used with learning rate = 0.001 and decay = 0.001. While learning rate acts as a gradient descent step size, decay parameter dampens this step size to take progressively smaller steps in each epoch. An $\ell_2$ loss function was minimized using 200 epochs and a batch size of 128. An NVIDIA Titan XP graphics card with 12 GB memory was used for training, which took ~18 hours for SAGE and ~19 hours for DWI.

For SAGE processing, slices and echoes were treated as different training instances. At the test stage, the patch-based network was applied in a sliding-window manner with a step size of 10 voxels, and the estimated residuals from overlapping patches were averaged together. This process took 4.5 seconds/slice. Similarly for DWI, slices and diffusion directions were treated as different training samples and the inference took 9 seconds/slice.

### Phase cycling

We form a magnitude estimate from the U-Net shot images by averaging, $m_{\text{unet}} = \frac{1}{N_s}\sum_{t=1}^{N_s}|u_t|$. We keep this improved magnitude fixed, and solve only for the image phase of each EPI-shot $\phi_t$ using phase-regularized parallel imaging, or phase cycling (24):

$$min_{\phi_t}\left\|F_t C m_{\text{unet}} \cdot e^{i\phi_t} - d_t\right\|_2^2 + \alpha\|W\phi_t\|_1 \qquad \text{Eq2}$$

Here, only the highlighted phase information $\phi_t$ is unknown, $W$ is a wavelet operator that imposes sparsity prior on the shot phase via $\ell_1$ penalty, and $\alpha$ is a parameter that controls the degree of regularization. The solution of this problem is made easier by the fact that we are using sensitivity encoding to solve only for the real-valued individual shot phases rather than the complex-valued individual shot images. The shot phases from the complex-valued SAGE network, $\sphericalangle u_t$, were used to initialize this non-convex problem. For DWI reconstruction with magnitude-based U-Net processing, shot phases from SMS-MUSSELS reconstruction, $\sphericalangle x_t$, served as initial guess. The complex-valued DWI network was still able to provide shot phase information, $\sphericalangle u_t$, to initialize phase cycling (Supporting Information Figs S9 and S10). SMS acceleration is again embedded in Eq2 via the 2-dimensional undersampling in $F_t$, and the coil sensitivities of the slices concatenated in the readout direction as represented by $C$.

### JVC-SENSE

Given estimates of shot-to-shot phase variations $\phi_t$, we can now jointly solve for the common magnitude image $m$ using the data from all shots, through harnessing sensitivity encoding for slice and in-plane acceleration (25) and the virtual coil (VC) concept (19,20) in JVC-SENSE. To do this, we solve a simple least squares problem:

$$min_m \sum_{t=1}^{N_s} \left\| \begin{bmatrix} F_t C e^{i\phi_t} \\ F_{-t} C^* e^{-i\phi_t} \end{bmatrix} m - \begin{bmatrix} d_t \\ d^*_{-t} \end{bmatrix} \right\|_2^2 + \beta \cdot \mathcal{R}(m) \qquad \text{Eq3}$$

Here, the only unknown is the highlighted magnitude image $m$, and the coil sensitivities are modified to include the phase variation in each shot to yield the combined sensitivities $C e^{i\phi_t}$. The VC concept is enforced by augmenting the optimization with the conjugate symmetric k-space data $d^*_{-t}$ and the conjugate sensitivities $C^* e^{-i\phi_t}$. Conjugate symmetric k-space is derived from the acquired k-space data by complex-conjugation and flipping the axes in the $k_x$-$k_y$ plane. We have used the shorthand notation $-t$ to express this mirroring operation in k-space. Joint reconstruction across all shots is performed via the summation operator $\sum_{t=1}^{N_s}(\cdot)$. For structural imaging with SAGE, we have used total variation (TV) penalty as the regularizer $\mathcal{R}(\cdot)$, with the corresponding regularization parameter $\beta$. For diffusion imaging, we have explored using TV regularization as well as a simple Tikhonov penalty.

During the review of this paper and after public dissemination of SMS-NEATR preprint (40), abstract (41) and code (https://bit.ly/2QgBg9U) which introduced SMS-MUSSELS and its improvement using deep learning, two preprints have also appeared that describe SMS-MUSSELS and an alternative deep learning enhanced msEPI reconstruction (42,43).

## Training Data

### Spin-and-gradient-echo (SAGE)

In compliance with Institutional Review Board (IRB) requirements, three volunteers were scanned on a Siemens Prisma 3T system with SAGE (22) msEPI sequence to build a training dataset. Multishot data were collected, where each shot was acquired at $R_{inplane}$=8-fold acceleration, and a total of 8-shots were collected with a $\Delta k_y$ sampling shift between the shots. When combined, this corresponded to a fully-encoded acquisition at $R_{net}$=1. Relevant parameters were: field of view (FOV) = 220x220x120 mm$^3$, resolution = 1x1x3mm$^3$, echo times (TEs) = 26/61/95/130/165 ms, repetition time (TR) = 8.3 sec, and effective echo spacing = 0.148 ms. Each shot sampled only 27 phase encoding lines due to $R_{inplane}$=8-fold acceleration. First two echoes were sampled before the 180° pulse, and the latter three were acquired after the refocusing pulse. The fifth echo was timed so that it was a spin echo image.

Coil sensitivities used in reconstructions were estimated using ESPIRiT (44,45) based on a FLEET acquisition (46). FLEET autocalibration signal (ACS) acquisition collects multishot gradient echo EPI data with low flip angles, where all the shots for a specific slice are acquired first. Then all shots for the second slice are sampled, and this is repeated until every slice in the FOV prescription are accounted for. This way, encoding of each slice is completed within a time frame on the order of 100 msec, and shot-to-shot variations are minimized. Unlike the FLEET calibration scan, the "standard mode" for ACS acquisition would sample the $1^{st}$ shot for all the slices first, then acquire the $2^{nd}$ shot again for all slice positions. The time frame for sampling all the shots is thus on the order seconds, which increases the vulnerability to shot-to-shot motion and has detrimental impact on the ACS data quality (46). Using FLEET acquisition has thus allowed us to improve the robustness of our coil sensitivity estimation. All acquisitions were made with a Siemens 32-channel head coil.

To obtain clean reference data, MUSSELS reconstruction was performed using all of the 8-shots at $R_{inplane}$=8 acceleration, which yielded "fully-sampled" ground-truth images. To enable higher acquisition efficiency, only 2-shots out of the 8-shot data were selected for subsampled reconstructions. The 2-shots were acquired with a k-space shift of $\Delta k_y$=4 samples to provide complementary coverage. These were further collapsed in the slice direction to simulate MB=2-fold acceleration, so that the total acceleration factor *per shot* became $R_{inplane}$ x MB =8x2. This highly undersampled msEPI data were then reconstructed using SMS-MUSSELS. Due to the very high acceleration rates, SMS-MUSSELS algorithm incurred reconstruction artifacts. These errors with respect to the clean reference image were used as the training target in our residual learning approach **(Fig. 2)**.

We extracted 57600 overlapping patches of size 64×64 with a step size of 16 voxels from the training data. The 2-shots decomposed into real and imaginary components in the SAGE acquisition were treated as input channels, and were concatenated to create 64×64×4 patches that were fed to the network to enable joint reconstruction across shots. The training dataset was enriched by 16-fold using augmentations including scaling (0.5×, 1×, 2×), flipping the axes (left-right, anterior-posterior and echo dimension) and rotations (±135, ±90, ±45 degrees).

### Diffusion Weighted Imaging (DWI)

Three volunteers were scanned on a 3T Prisma system to build up a DWI training dataset, consisting of 9-shot data acquired at $R_{inplane}$=9-fold acceleration. The parameters were: field of view (FOV) = 224x224x120 $mm^3$, resolution = $1x1x3mm^3$, TE/TR = 54/5100 ms, and effective echo spacing = 0.13 ms. Each shot sampled only 24 phase encoding lines due to $R_{inplane}$=9-fold acceleration. In addition to a b=0 image, six diffusion directions at b=1000s/$mm^2$ were collected. FLEET calibration data were used to estimate ESPIRiT coil sensitivity maps.

Reference "fully-sampled" images were obtained using all 9-shots in MUSSELS reconstruction. Subsampled acquisitions were obtained by selecting 5-shots out of this 9-shot dataset. The 5-shots were shifted by $\Delta k_y$={0,2,4,6,8} samples to provide complementary information. These were further undersampled by collapsing two slices that are 60 mm apart to simulate MB=2 slice acceleration. The highly subsampled diffusion msEPI data (R$_{inplane}$xMB=9x2) were reconstructed using SMS-MUSSELS. Reconstruction errors with respect to the "fully-sampled" reference data were learned using a deep network. Similar patch extraction and augmentation steps were performed.

## Reconstruction Experiments

**SMS-MUSSELS parameter optimization:** We explored the dependence of the reconstruction performance of SMS-MUSSELS on the k-space window size $r$, as well as the rank constraint enforced by the number of singular values, $k$. For the SAGE dataset, we evaluated the RMSE metric on a slice group from a training subject, and considered a range of window sizes $r \in \{2,3,4,5,6,7\}$. To control the rank of the data matrix $\mathcal{H}(x)$ which has $N_s \times r \times r$ columns in an intuitive manner, we varied the "effective number of shots $(N_{eff})$" between $N_{eff} \in \{0.75, 1, 1.25, 1.5\}$. For instance, using a window size $r = 6$ and enforcing $N_{eff} = 1.25$ would imply that the number of singular values $k = N_{eff} \times r \times r = 45$ is used during the reconstruction. This way, $N_{eff}$ gives us a handle on the rank constraint in terms of the effective number of shots we allow the msEPI data to have. The optimal parameter setting turned out to be $r = 5$ and $N_{eff} = 1$ for the 2-shot SAGE reconstruction at R$_{inplane}$xMB=8x2. Termination criterion was less than 0.1% update between image estimates from successive iterations. This analysis is presented in Supporting Information Figure S1.

Using DWI dataset from a training subject, a similar analysis revealed that the optimal parameter setting is $r = 7$ and $N_{eff} = 1.25$ for a 5-shot reconstruction at R$_{inplane}$xMB=9x2. Best RMSE was obtained by terminating the SMS-MUSSELS iterations when the update in the image estimate between iterations was less than 0.3%.

**POCS versus FISTA updates:** To improve the convergence rate and image quality of our POCS-like optimization algorithm for SMS-MUSSELS, we have explored FISTA update rule, which makes use of a combination of the current and previous iterates to form the next image estimate (as detailed in the Appendix). FISTA was previously used in the context of diffusion msEPI with local low rank constraint in image-space (47). A comparison on one of the SAGE training datasets indicated that FISTA provided substantial reduction of aliasing/ghosting artifacts as well as RMSE improvement, when other parameters were held constant ($r = 5$ and $N_{eff} = 1$). As such, we have used FISTA iterations for the remaining SMS-MUSSELS reconstructions reported herein. Convergence analysis is provided in Supporting Information Fig S11.

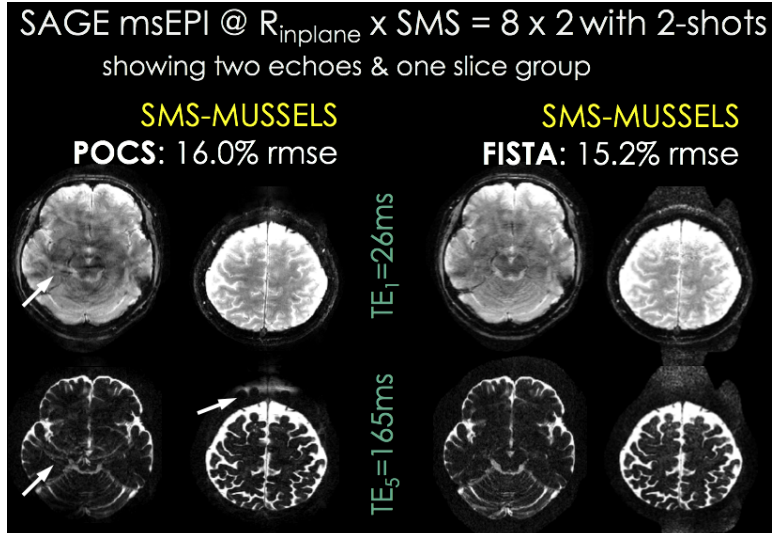

*Figure 3 $R_{inplane}$xMB=8x2-fold accelerated SAGE msEPI acquisition with 2-shots from a training dataset. One SMS slice group and two echoes out of a total of five are depicted. Using a POCS-like solver for SMS-MUSSELS optimization led to residual aliasing/ghosting artifacts (arrows). FISTA update rules improved convergence and image quality of SMS-MUSSELS, and mitigated these structured errors*

**SAGE reconstruction @ $R_{inplane}$xMB=8x2 with 2-shots:** msEPI SAGE data were acquired on a 4[th] subject (not seen during the training of the network). This acquisition was then reconstructed with SMS-MUSSELS using FISTA iterations and the optimized parameter setting in Matlab, running on a workstation with 64 CPU processors and 256 GB memory.

SMS-MUSSELS shot images were then processed with the trained U-Net, which allowed improved estimation of shot-phases using phase cycling. We used 500 iterations and "db4" wavelets in phase cycling, and set the regularization parameter to $\alpha=10^{-5}$ for optimal RMSE. Having estimated the phase of each shot, JVC-SENSE with total variation penalty ($\beta=3\cdot10^{-4}$) was used to compute the final magnitude image. For all the remaining experiments, we used these reported parameter values without further optimization.

**SAGE reconstruction *without* ML:** To assess the contribution of the ML step to SMS-NEATR, we performed an additional reconstruction *without* U-Net processing. We used the same undersampling setup from the first experiment, namely $R_{inplane}$xMB=8x2 acceleration with 2-shots. Starting from the SMS-MUSSELS magnitude estimate $m_{\text{mussels}}$, we employed phase cycling to solve:

$$min_{\phi_t} \| F_t C m_{\text{mussels}} \cdot e^{i\phi_t} - d_t \|_2^2 + \alpha \| W \phi_t \|_1 \qquad \text{Eq4}$$

Having obtained refined shot-to-shot phase estimates $\phi_t$, we went on to use JVC-SENSE and arrive at a refined magnitude solution. This way, we followed the flowchart outlined in Fig. 1, except that we by-passed the U-Net processing step.

**SAGE reconstruction *using BM3D instead of U-Net*:** We have also explored replacing the deep network with a conventional denoiser, BM3D (48), to help improve the SMS-MUSSELS output. After decomposing the shot images into real and imaginary components, we processed each of these images separately, and normalized their intensity to be within [0,1]. We optimized the BM3D filter width $\sigma$ for the best RMSE. The resulting shot images were used to initialize phase-cycling and JVC-SENSE reconstructions, i.e. we again followed the flowchart in Fig. 1, but replaced U-Net with BM3D.

**$T_2$ and $T_2^*$ parameter fitting using SAGE data:** The five echo images produced by SMS-MUSSELS and SMS-NEATR algorithms were used in a Bloch-equation based model fit (22) to estimate $T_2$ and $T_2^*$ parameter maps. As supplementary information, we have also explored parameter fitting to the reconstructions obtained from U-Net and BM3D denoisers.

**DWI reconstruction @ $R_{inplane}$xMB=9x2 with 5-shots:** DWI data at six directions were acquired on a 4$^{th}$ subject (not seen during the training). Two of these directions were reconstructed with SMS-MUSSELS. These images were refined using the deep diffusion network, and were processed with phase-cycling and JVC-SENSE to compute SMS-NEATR results. In this case, 50 phase-cycling iterations with $\alpha=10^{-3}$, and Tikhonov regularized JVC-SENSE with $\beta=10^{-2}$ yielded optimal RMSEs. We have also explored BM3D denoising, again with complex-valued processing for each shot separately, and optimized for the filter width $\sigma$.

**DWI analysis:** Six direction diffusion data reconstructed by the algorithms under comparison were registered using *MCFLIRT* (49). Diffusion tensor fitting was performed using the *DTIFIT* function in FSL, which also produced fractional anisotropy (FA) and mean diffusivity (MD) maps.

## RESULTS

**SAGE reconstruction @ $R_{inplane}$xMB=8x2 with 2-shots:** The left column of Fig. 4 shows SMS-MUSSELS reconstructions, where only the first and last echoes and root-sum-of-squares error map calculated over the entire 5 echoes are displayed. U-Net processing mitigated some of the noise amplification and improved the RMSE from 10.8% to 8.3%. Starting from this, SMS-NEATR was able to provide a small error reduction (8.1%), with similarly high image quality.

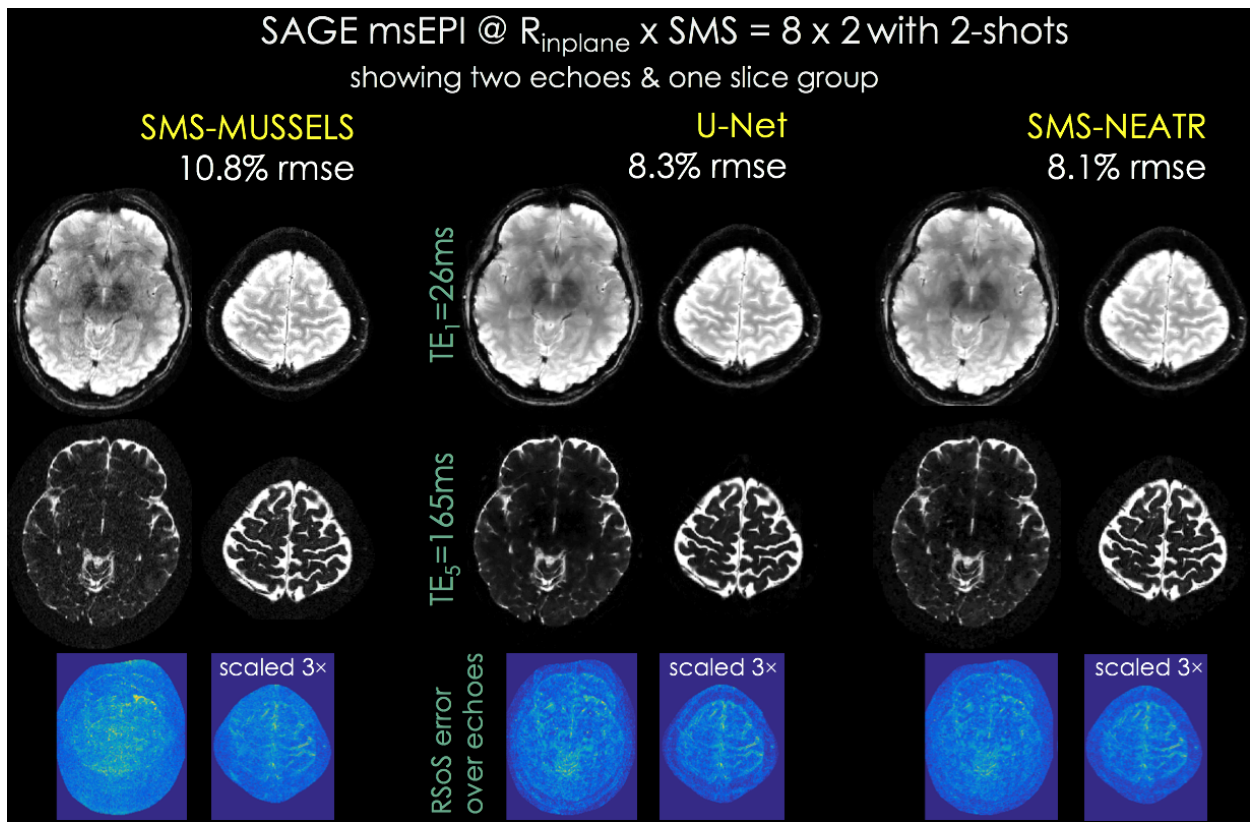

*Figure 4 SAGE test dataset at $R_{inplane}$xMB=8x2-fold acceleration using 2-shots. The first and last echoes are shown for a single SMS slice group. SMS-MUSSELS with FISTA (left) was successful in reconstructing images despite the high acceleration with 10.8% error. The bottom row shows root-sum-of-squares combination of error images across the five echoes. U-Net denoising of SMS-MUSSELS reconstruction provided improvement (8.3%, middle), and was used for initializing SMS-NEATR for additional quality gain (8.1%, right).*

**SAGE reconstruction *without* ML:** Supporting Information Fig S2 demonstrates the effect of not using U-Net denoising in the SMS-NEATR pipeline. In this case, there was still some gain from refining the shot-phase estimates using phase-cycling and joint parallel imaging reconstruction (RMSE went from 10.8% to 9.2%), but the improvement over SMS-MUSSELS was yet higher when ML was included (8.1%).

**SAGE reconstruction *using BM3D instead of U-Net*:** Using a conventional BM3D denoiser could still provide RMSE reduction (9.3%) but the learned U-Net model was more successful in refining the SMS-MUSSELS output (8.3%). Supporting Information Fig S3 also explores using BM3D to replace U-Net in the SMS-NEATR flowchart, which appeared to be slightly less effective (8.4% with BM3D initialization versus 8.1% with U-Net jumpstart).

**$T_2$ and $T_2^*$ parameter fitting using SAGE data:** Parameter maps from a slice group reconstructed using SMS-MUSSELS and SMS-NEATR are depicted in Fig. 4, corresponding to an 8.3sec acquisition with whole-brain coverage at 1x1x3mm$^3$ resolution. SMS-NEATR was able to mitigate noise amplification and image artifacts mainly affecting the middle of the FOV in the SMS-MUSSELS maps. $T_2$ and $T_2^*$ fits after BM3D and U-Net denoising are compared in Supporting Information Fig S4. U-Net estimates again appeared to have higher quality than the BM3D results, and were similar to the SMS-NEATR maps.

Supporting Information Figs S12 and S13 show parameter maps from fully-sampled MUSSELS reconstruction as well as error maps from the accelerated reconstructions.

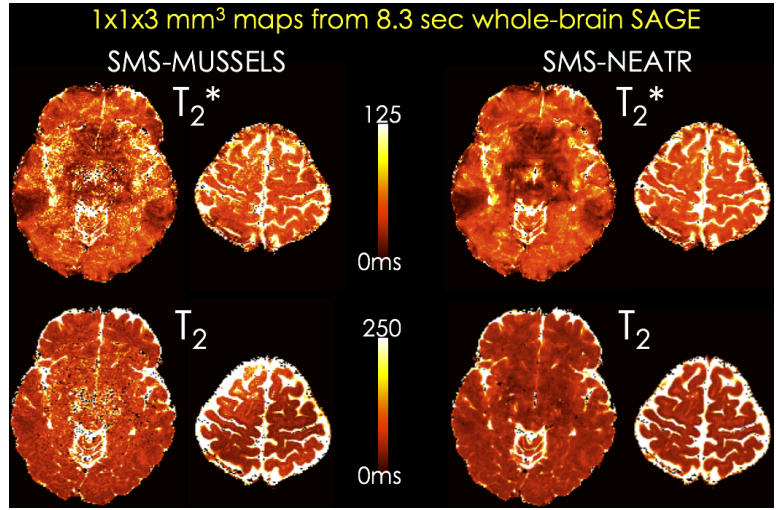

*Figure 5* $T_2$ and $T_2^*$ parameter maps obtained by Bloch equation fitting to the five-echo SAGE reconstruction. This 2-shot acquisition at $R_{inplane}$xMB=8x2-fold acceleration provides whole-brain coverage in 8.3 sec with low geometric distortion. While SMS-MUSSELS parameter maps appeared noisy (left), these artifacts were mitigated in the SMS-NEATR estimates (right).

**DWI reconstruction @ $R_{inplane}$xMB=9x2 with 5-shots:** Fig. 6 shows DWI slice groups from one diffusion direction. These lower slices with poor $B_0$ uniformity were selected to demonstrate the effect of high in-plane acceleration in avoiding distortion and voxel pile-up artifacts. Regardless, SMS-MUSSELS did suffer from residual ghosting and noise amplification in these difficult reconstruction tasks. BM3D and U-Net processing helped denoise the data, but failed to eliminate the structured artifacts (indicated by white arrows). BM3D results appeared over-smooth, whereas U-Net provided a better trade-off between over-smoothing and denoising. SMS-NEATR did not suffer from over-smoothing, and could mitigate noise amplification and structured artifacts. We anticipate that RMSE values have contributions from both noise and reconstruction error since the ground truth data are also noisy. As such, RMSE is likely to be partially indicative of reconstruction performance (U-Net consistently had the best performance).

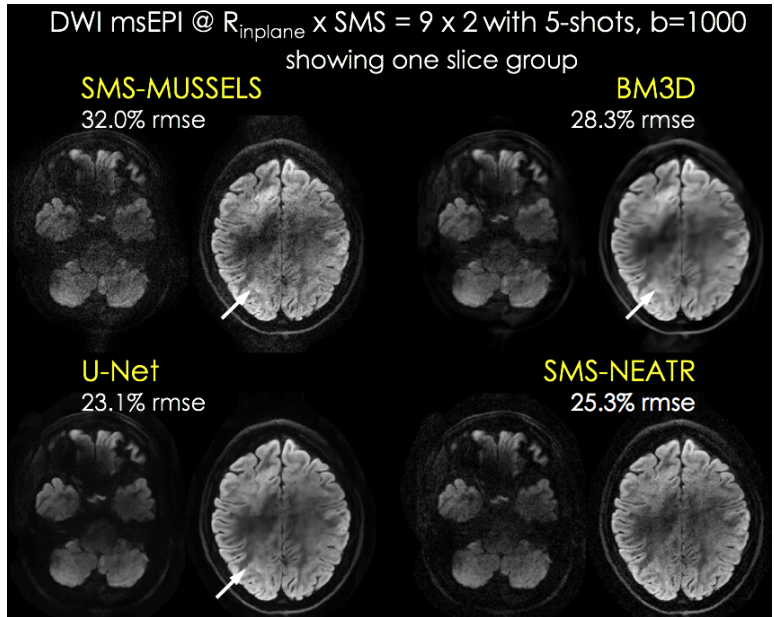

*Figure 6 An SMS slice group of a second diffusion direction from the test msEPI acquisition is shown. SMS-MUSSELS suffered from noise amplification and some structured artifacts. BM3D and U-Net could denoise the SMS-MUSSELS result at the potential cost of over-smoothing, while some artifacts persisted (arrows). SMS-NEATR could provide better SNR and image quality without the vulnerability to over-smoothing.*

Using complex-valued or magnitude-based U-Net processing led to similar SMS-NEATR results in DWI (Supporting Information Fig S10). TV-regularizer could provide further RMSE reduction than Tikhonov penalty, but this came at the cost of some over-smoothing (Supporting Information Fig S8). Reducing the TV regularization parameter led to comparable RMSEs and image sharpness as $\ell_2$-penalty using both complex- and magnitude-valued U-Net initialization (Supporting Information Fig S8 and S9).

Fig. 7 shows SMS-NEATR results from the six direction acquisition, as well as the average DWI image, color FA and MD maps, and the root-sum-of-square combination of error images across all directions. A similar analysis is presented for the fully-sampled, SMS-MUSSELS and U-Net reconstructions in Supporting Information Figs S14 – S16.

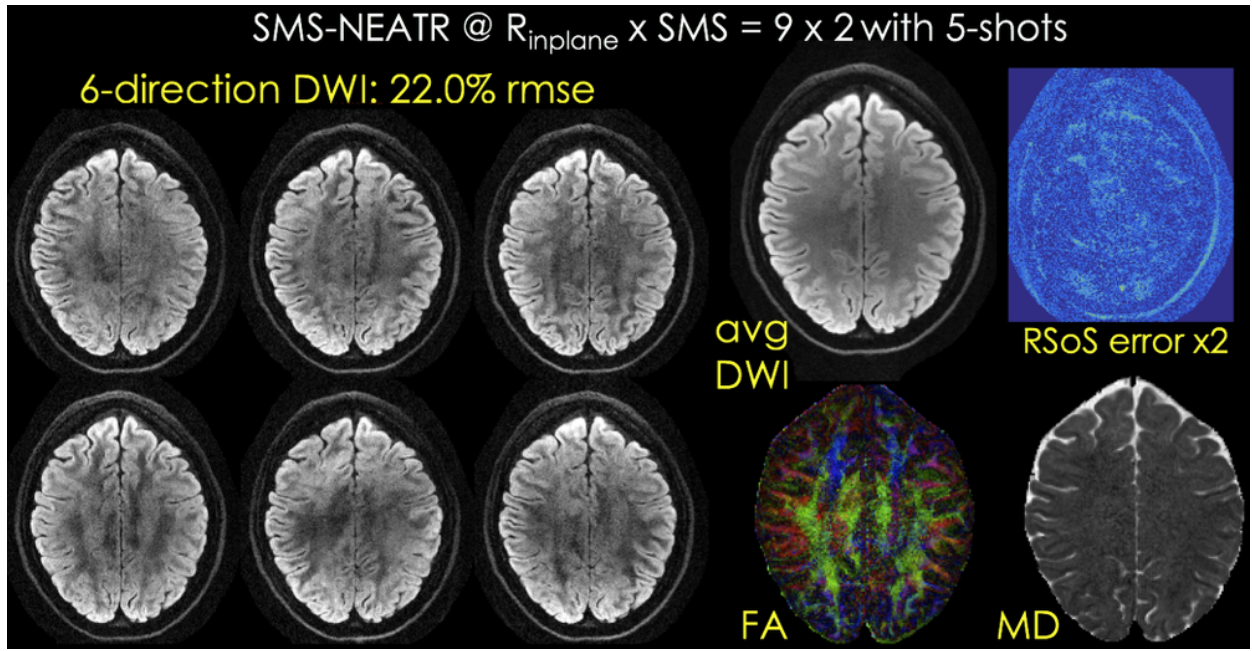

*Figure 7* SMS-NEATR reconstruction for six direction diffusion data, as well as average DWI, color FA and MD maps and root-sum-of-squares error across the directions are presented.

## DISCUSSION

We presented SMS-NEATR, a synergistic ML and physics-based reconstruction approach, that allowed up to $R_{net}$=8-fold accelerated msEPI with high image quality. This was made possible by taking advantage of phase-cycling algorithm, the newly developed SMS-MUSSELS, deep learning, and joint parallel imaging reconstruction. Our residual CNN learned to predict and mitigate the errors in highly accelerated SMS-MUSSELS reconstruction, which then permitted phase-cycling to estimate shot-to-shot phase variations. Including this information as additional sensitivity variations then allowed JVC-SENSE to solve for a common magnitude image using the entire multishot k-space data and VC concept.

Partial Fourier sampling was *not* performed during any of the acquisitions. While this would have helped achieve shorter TE and higher SNR in DWI, it would not affect the geometric fidelity of the acquisition. Our motivation in employing high in-plane acceleration rates was to reduce distortion and $T_2^*$-related blurring, as well as enabling shorter TE. Particularly with the SAGE scan, our aim was to replace the currently inefficient spin-warp imaging with the much faster, msEPI-based acquisitions for rapid clinical imaging, while minimizing ghosting, blurring and distortion artifacts that plague EPI.

CNNs can represent very complicated and non-linear input/output relations. While this makes them very powerful, such a complex mapping between input and output causes the network to be difficult to characterize. Since its direct application may lead to unpredictable errors, end-to-end CNN reconstruction in clinical settings is likely to raise acceptance issues. A ML reconstruction approach that can overcome this issue was proposed in the Variational Network (VN) (50) formulation. This allows a transparent deep learning reconstruction of accelerated acquisition where both the kernel weights and nonlinear activation functions are learnt and can be visualized at any layer. VN also utilizes sensitivity encoding and enforces consistency to the acquired k-space data. Similarly, model based deep learning (MoDL) is powerful in its ability to combine data consistency and convolutional layers (51). These ideas treat the iterations in gradient-descent type reconstructions as unrolled networks to retain fidelity to acquired data via a forward model, while learning model parameters that map the reconstruction to a reference image (52). Importantly, such combination of a forward-model and learned filtering provided further improvement than a model-based reconstruction followed by U-Net denoising (53).

SMS-NEATR also taps into the potential of CNN without treating it as an end-to-end tool, while fully harnessing the encoding provided by the scanner hardware. We achieved this by using CNN to obtain an interim image with minimal artifacts, while utilizing a rigorous physics-based approach to validate and improve upon this solution in the final step of reconstruction. Our goal in SMS-NEATR is to capture shot-to-shot phase variations accurately, since when they are known, a JVC-SENSE reconstruction that solves for the magnitude image is capable of outperforming alternative approaches. Synergistic combination of ML and physics-based reconstruction proved to be powerful, leading to ~30% RMSE reduction over our SMS-MUSSELS implementation (**Figs. 4 and 6**). In the absence of deep learning initialization, the subsequent phase cycling and JVC-SENSE steps provided a smaller, <20% improvement over the SMS-MUSSELS reconstruction (Supporting Information Fig. S2). A conventional BM3D denoiser also proved to be effective in jumpstarting SMS-NEATR, but the performance was consistently better using a learned denoiser tailored for the specific application (**Figs. 6**, Supporting Information Figs. S3 and S4). We anticipate further gains from advanced models that could simultaneously enforce data consistency and perform learned filtering (54–56). This would also streamline the SMS-NEATR pipeline and reduce the number of steps.

Application of msEPI in structural imaging is made difficult by ghosting artifacts from hard-to-estimate physiologic signal changes between shots. This is particularly true for gradient echo imaging at late TEs. To illustrate, we performed a "sliding window" combination of 8-shots of SAGE data acquired at $R_{inplane}$=8 acceleration to obtain "fully-sampled" data (Supporting Information Fig. S5). The ghosting artifacts that stem from physiological noise is especially strong in the 2$^{nd}$ and 3$^{rd}$ echoes due to increased phase accrual at long TEs (the last echo is in fact a spin echo, which refocuses most of the phase evolution and results in the cleanest image). Using a standard forward-model based reconstruction for structural msEPI would thus necessitate the simultaneous estimation of the image content and the phase variations in each shot. Since both the clean image and the phase information used in the forward-model are unknown, this would entail the solution of a computationally prohibitive, non-convex optimization problem that could get stuck at local minima. As such, existing msEPI techniques circumvent this difficulty by dividing the reconstruction into two separate parts: shot-phase estimation, and combination of multishot given the estimated phase information. Navigator-based approaches derive this phase information from additional calibration acquisitions made for each shot (3–7). Diffusion imaging with MUSE and its extensions (9,12,13) operate without a navigator, and perform the phase estimation step using parallel imaging to reconstruct a complex image for each shot. Smoothing the phase of each intermediate image then yields an estimate of shot-to-shot variations, which allows joint reconstruction of all multishot data together.

msEPI reconstruction is indeed harder in diffusion imaging, since the phase variations amplified by diffusion gradients can be much stronger than the physiologic noise in structural imaging, as demonstrated in the "sliding window" combination in Supporting Information Fig. S6. The final Supporting Information Fig. S7 shows a similar sliding window data combination, but this time without any diffusion gradients (b=0). Even in this case where one would not expect any artifacts, there are minor ghosts that may be stemming from patient motion. Given that the TR was 5.1sec, msEPI acquisitions are indeed susceptible to motion artifacts since it took ~46sec to sample these 9-shots. A side benefit of highly accelerated msEPI could be an improvement in motion robustness. We have seen similar gains in the final SMS-NEATR diffusion reconstructions using either magnitude- or complex-valued deep learning. We expect this was because the magnitude network could provide higher quality magnitude priors which helped phase-cycling to better solve for the shot-phase data, whereas the complex network provided an overall gain in both magnitude and phase estimates – but the magnitude output was improved to a lesser extent (as can be seen in the RMSE values in Supporting Information Fig S10). Having obtained similar SMS-NEATR results may indicate that there is flexibility in the blocks in the pipeline, as long as the shot-phase estimates are improved beyond those of SMS-MUSSELS.

MUSSELS exploits similarities between the shot-images using a low-rank prior on the block-Hankel representation of their k-space (11,14), so that it can perform msEPI reconstruction without explicit shot-phase estimation. MUSSELS has allowed $R_{inplane}$=8-fold acceleration per each shot in msEPI diffusion imaging using as few as 4-shots ($R_{net}$=2). Unlike earlier navigator-based (5–7) or navigator-free approaches (8–10) where the number of acquired shots was equal to the in-plane acceleration factor ($N_s$=$R_{inplane}$), MUSSELS could thus perform in the ($N_s$<$R_{inplane}$) regime to improve acquisition efficiency. With SMS-NEATR, we pushed the efficiency gain even further to enable $R_{net}$=8-fold acceleration ($R_{inplane}$xMB=8x2 with 2-shots) in structural imaging, and $R_{net}$=3.6-fold ($R_{inplane}$xMB=9x2 with 5-shots) in diffusion imaging. Although SMS-MUSSELS had some residual artifacts at such high accelerations, it provided a good initial guess for our residual network to further clean up the shot-images. Starting from these estimates, we could then solve for the phase variations using phase cycling, which constituted an easier problem since the unknown information was a real-valued phase image. This provided a 2-fold reduction in the number of unknowns compared to a complex-valued SENSE solution.

The best RMSE performance for structural imaging was obtained with $r \times r$ = 5×5 windows and with $N_{eff}$=1, whereas the optimal parameters were $r \times r$ = 7×7 windows and with $N_{eff}$=1.25 for DWI. The increased window size and rank constraint should help capture greater shot-to-shot variations, which are more likely to be observed in diffusion imaging than the SAGE scan. Further relaxing the rank constraint and using larger windows could help represent more spatially varying phases differences between the shots, but relaxing these priors beyond their optimal values would come at the potential cost of RMSE performance. Indeed, using $N_{eff} = N_s$ would be a non-informative prior and the outcome would be identical to a shot-by-shot SENSE reconstruction.

**Limitations and their mitigation:** SMS-NEATR uses ML to provide an initial estimate to a difficult image reconstruction problem, thereby avoiding the vulnerability of poor generalization of "direct" ML reconstruction. In addition to this, we have augmented our training dataset size by 16-fold to subject the network to greater variation. Using a patch-wise representation with overlapping patches helped further increase the available number of training samples. Finally, we have provided the network with different contrasts (echoes or diffusion directions) as training samples to help improve generalization. Despite these precautions, the network would benefit from re-training if large changes in sequence parameters are desired to be made, or if they are dictated by hardware limitations of other scanners. We also anticipate that having the subsequent physics-based reconstruction will mitigate some of the generalization concerns, as the ML output is used for initializing this model based step. Exploring unrolled networks with data consistency layers (50–53,55) or using conventional denoisers could provide additional robustness. Using smoothness priors embedded in the MUSSELS reconstruction with the SR-MUSSELS formalism rather than relying on learned or conventional denoisers would also be an elegant solution.

Another consideration is the selection of the reconstruction technique that provides the initial solution to U-Net. We have developed a FISTA-based solver for SMS-MUSSELS, but other advanced reconstruction strategies such as MUSE (9) or POCS-MUSE (10) could also be utilized to provide this initial estimate.

*Qualitatively*, SMS-NEATR provided greater gains in the more challenging multishot DWI reconstruction than the SAGE application. It has better mitigated ghosting/aliasing artifacts and noise amplification than SMS-MUSSELS, but the RMSE metrics remained above 20%. We think that this is because the ground truth diffusion data is also corrupted by noise, which makes it difficult to disentangle reconstruction artifacts from the noise contribution. As such, other measures of fidelity to reference data could better gauge the improvement in DWI reconstruction.

For Nyquist ghost correction, we have used a simple 1-dimensional navigator in a slice-specific manner. Especially in oblique acquisitions, more involved ghost correction techniques such as Dual Polarity Grappa (57) and LORAKS (31) should allow for improved suppression of these errors. To ensure that the residual ghosts seen in presented results stem only from reconstruction errors due to acceleration, we have included fully-sampled MUSSELS reconstructions and FLEET calibration data that do not exhibit ghosting in the Supporting Information Figs S17 and S18.

**Extensions:** We have demonstrated the applications of SMS-NEATR in msEPI SAGE and DWI acquisitions. Enabling $R_{inplane}$=8 or 9-fold acceleration in other pulse sequences could help create a multi-contrast msEPI clinical protocol with high geometric fidelity. This would minimize the distortion and blurring artifacts that hamper image quality and achievable resolution in the recently developed singleshot EPI protocols (1,2). Employing msEPI readout in multi-inversion $T_1$ mapping (58,59) and FLAIR (60) acquisitions with SMS-NEATR reconstruction could enable a rapid MR exam with similar table time as a CT scan. Other advanced encoding strategies such as wave-EPI (61) could provide additional efficiency gain and/or in-plane acceleration capability.

We believe that the strategy of utilizing ML to estimate unknown nuisance parameters in physics-based forward model reconstructions can be impactful in solving other prohibitively difficult problems. We have recently demonstrated this concept in prospective motion correction (62), where we used residual deep learning to provide an interim image with largely reduced motion artifacts. This interim CNN reconstruction provides an initial image and motion parameter estimate thus jumpstarting the physics-based TAMER algorithm (63), which uses the extra degrees of freedom in multi-coil data to jointly estimate motion parameters and the clean image. Having access to a good initial guess helped the non-convex TAMER optimization converge 30× faster to the final solution. Other venues that might benefit from this synergistic approach could be in navigator-free Nyquist (N/2) ghost correction, calibrationless parallel imaging and reference-free k-space trajectory estimation.

## CONCLUSION

We demonstrated the ability of SMS-NEATR, a combined ML and physics-based reconstruction algorithm, in providing high quality reconstructions from up to 8-fold accelerated msEPI acquisitions using 2–5 shots of data. The ability to acquire high in-plane resolution images with minimal distortion and blurring could enable an msEPI-based MRI exam with multiple contrasts, while matching the table time of a CT scan.

## ACKNOWLEDGMENT

We acknowledge a GPU grant from NVIDIA, and support from NIH NINDS (K23 NS096056) NIMH (R24 MH106096 and R01 MH116173) and NIBIB (U01 EB025162, R01 EB020613, R01 EB019437, R01 EB017337 and P41 EB015896). Additional support was provided by the MGH/HST Athinoula A. Martinos Center for Biomedical Imaging. This research was made possible by the resources provided by NIH Shared Instrumentation Grants S10-RR023401 and S10-RR023043.

## APPENDIX

We pursue a POCS-like solution to the optimization problem posed in the MUSSELS formalism (Eq1), and follow the steps below:

```
y₁ = x₀     % initial guess from e.g. SMS-SENSE reconstruction
τ₁ = 1
for i = 1: N_iter
    % Low-rank constraint:
    A = H(y_i)
    UΣV^H = svd(A)
    A = UΣ_k V^H     %Σ_k is obtained via hard thresholding by keeping the k largest singular values
    x_i = H*(A)     %H* is a transposed mapping that inserts Hankel matrix elements into multi-shot k-space

    for t = 1: N_s
        x_t = x_i(:,:,t)
        % Generate coil images by multiplication with sensitivities:
        x_c = C x_t
        % Resubstitute acquired k-space:
        x_c = x_c + F^H (d_t − F_t x_c)
        % Coil combination:
        x_t = (C^H C)^(−1) C^H x_c
        x_i(:,:,t) = x_t
    end
    if use_fista
        τ_{i+1} = (1 + √(1 + 4τ_i²)) / 2
        y_{i+1} = x_i + ((τ_i − 1)/τ_{i+1})(x_i − x_{i−1})
    else
        y_{i+1} = x_i
    end
end
```

The flag "*use_fista*" toggles between conventional POCS-like update rule and FISTA iteration, which has earlier iterates to form the next image estimate. $N_{iter}$ denotes the maximum number of iterations, which we have taken to be 200. $x_0$ is an initial guess for the msEPI images, and were estimated using an SMS-SENSE reconstruction for each of the shots independently.

**FIGURE CAPTIONS**

*Fig 1.* SMS-NEATR is a combined machine learning and physics-based reconstruction technique for highly-accelerated msEPI acquisition. We developed SMS-MUSSELS algorithm to provide an initial solution, which may suffer from artifacts due to high acceleration ($R_{inplane}$xMB=8x2 with 2-shots). Starting from this, residual learning with U-Net architecture provides an interim image with minimal artifacts. Given this solution, phase cycling algorithm is used for estimating shot-phases, which are then utilized as sensitivity variations in a final joint virtual coil (JVC) SENSE reconstruction.

*Fig 2.* U-Net architecture is used to learn the mapping between patches of shot-images reconstructed with SMS-MUSSELS, and their difference to reference data. Both the input and output have been decomposed into real and imaginary components to enable complex-valued processing for SAGE reconstruction. 64×64 patches from all the shots are presented as input to a 5-level network, where the first level uses 64 convolutional filters. To help provide scale invariance, max pooling operators downsample the patches after each layer. At the same time, the number of filters are doubled to retain the total number of kernel weights at each level.

*Fig 3.* $R_{inplane}$xMB=8x2-fold accelerated SAGE msEPI acquisition with 2-shots from a training dataset. One SMS slice group and two echoes out of a total of five are depicted. Using a POCS-like solver for SMS-MUSSELS optimization led to residual aliasing/ghosting artifacts (arrows). FISTA update rules improved convergence and image quality of SMS-MUSSELS, and mitigated these structured errors.

*Fig 4.* SAGE test dataset at $R_{inplane}$xMB=8x2-fold acceleration using 2-shots. The first and last echoes are shown for a single SMS slice group. SMS-MUSSELS with FISTA (left) was successful in reconstructing images despite the high acceleration with 10.8% error. The bottom row shows root-sum-of-squares combination of error images across the five echoes. U-Net denoising of SMS-MUSSELS reconstruction provided improvement (8.3%, middle), and was used for initializing SMS-NEATR for additional quality gain (8.1%, right).

*Fig 5.* $T_2$ and $T_2^*$ parameter maps obtained by Bloch equation fitting to the five-echo SAGE reconstruction. This 2-shot acquisition at $R_{inplane}$xMB=8x2-fold acceleration provides whole-brain coverage in 8.3 sec with low geometric distortion. While SMS-MUSSELS parameter maps appeared noisy (left), these artifacts were mitigated in the SMS-NEATR estimates (right).

*Fig 6.* Diffusion msEPI acquisition at $R_{inplane}$xMB=9x2 acceleration with 5-shots from the test subject. One SMS slice group is shown for this whole-brain acquisition. SMS-MUSSELS suffered from aliasing/ghosting artifacts in this harder reconstruction problem. BM3D and U-Net denoising could mitigate noise, but the structured artifacts persisted (arrows). SMS-NEATR was able to further mitigate these errors to improve image quality, while avoiding potential over-smoothing BM3D and U-Net may suffer from.

*Fig 7.* SMS-NEATR reconstruction for six direction diffusion data, as well as average DWI, color FA and MD maps and root-sum-of-squares error across the directions are presented.

SUPPORTING INFORMATION FIGURE CAPTIONS

*Supporting Information Fig S1*. Dependence of the reconstruction performance of SMS-MUSSELS on the k-space window size $r$, and the rank constraint as represented by the effective number of shots ($N_{eff}$) for a slice group from a SAGE training dataset. The optimal parameter setting was $r=5$ and $N_{eff}=1$ for the 2-shot SAGE reconstruction at $R_{inplane}$xMB=8x2 acceleration.

*Supporting Information Fig S2.* SAGE msEPI reconstruction results from the test dataset at $R_{inplane}$xMB=8x2 acceleration using 2-shots. The first and last echoes are displayed out of a total of five echoes belonging to this SMS slice group. SMS-MUSSELS yielded 10.8% RMSE (left), and was also used to initialize phase-cycling and JVC-SENSE reconstruction without machine learning (9.2% error, middle). Using U-Net to refine the SMS-MUSSELS result and jumpstart SMS-NEATR provided further improvement at 8.1% RMSE.

*Supporting Information Fig S3*. The SMS-MUSSELS reconstruction for the slice group in Fig S2 was denoised using BM3D and U-Net, where deep learning proved to be advantageous (BM3D: 9.3% versus U-Net: 8.3% error). Utilizing each of these denoised outputs to jumpstart SMS-NEATR led to similar reconstructions, and U-Net initialization had the best overall performance (8.1% RMSE).

*Supporting Information Fig S4*. Bloch equation based signal modeling for the five echoes in the SAGE acquisition allows for $T_2$ and $T_2$* parameter mapping. Denoising the 2-shot SMS-MUSSELS reconstruction at $R_{inplane}$xMB=8x2 acceleration, corresponding to an 8.3sec whole-brain acquisition, using BM3D and U-Net led to improvements in the quality of these quantitative maps. U-Net appeared more successful in mitigating the noise amplification in the middle of the FOV than the conventional BM3D filtering.

*Supporting Information Fig S5*. Sliding window (summation across shots in k-space) combination of 8-shots of SAGE data acquired at $R_{inplane}$=8 acceleration per shot. Due to physiological shot-to-shot phase variations, the combined fully-encoded images exhibit ghosting artifacts. This becomes more severe at later TEs, but is mitigated at the last echo, which is a spin echo image. The bottom row is scaled up to better demonstrate the artifacts.

*Supporting Information Fig S6*. Sliding window combination of 9-shots of DWI data acquired at $R_{inplane}$=9 acceleration per shot. Five slices from a whole-brain acquisition are depicted. Due to shot-to-shot phase variations stemming from motion under the diffusion encoding gradients, there are severe ghosting artifacts in these otherwise fully-encoded images.

*Supporting Information Fig S7*. Sliding window combination of data from the same acquisition session as Fig S6, but this time the diffusion gradients have been switched off (b=0). These spin echo images look relatively devoid of ghosting artifacts, but the scaled-up images in the bottom row reveal that these errors persist. We anticipate that head motion contributed to these artifacts, since it took around 46 sec to sample 9-shots at TR=5.1 sec.

*Supporting Information Fig S8.* Employing TV-regularization in JVC-SENSE led to similar results as L2 penalty. Reconstruction obtained with the TV parameter value that yielded the optimal RMSE value ($\lambda_{TV} = 10^{-2}$) appeared over-smooth, hence reducing the regularization to $\lambda_{TV} = 3 \cdot 10^{-3}$ provided a better trade-off between RMSE performance and image sharpness. Magnitude-based U-Net was used to initialize SMS-NEATR in these reconstructions.

*Supporting Information Fig S9.* Using complex-valued deep learning to initialize JVC-SENSE yielded similar quality reconstructions as magnitude-based U-Net processing. JVC-SENSE could flexibly utilize L2 or TV regularizers to further stabilize the reconstruction with comparable results.

*Supporting Information Fig S10.* Both complex- and magnitude-valued U-Net processing could denoise diffusion images, albeit at the cost of remaining artifacts (arrows) and especially in the case of magnitude-valued network, over-smoothing. Using these to jumpstart SMS-NEATR led to crisp images with mitigated artifacts.

*Supporting Information Fig S11.* FISTA update rule helped stabilize SMS-MUSSELS reconstruction, especially in later iterations where POCS still experienced large signal updates. Different colors indicate different echo images. Termination criterion was reaching less than 0.1% change between successive image estimates and maximum iteration number was 200.

*Supporting Information Fig S12.* Parameter maps from fully-sampled MUSSELS reconstruction, corresponding to an 8-shot, 66.4 second SAGE scan.

*Supporting Information Fig S13.* Parameter error maps from $R_{inplane}$ x SMS = 8 x 2-fold accelerated SMS-MUSSELS, U-Net and SMS-NEATR reconstructions relative to the fully-sampled data.

*Supporting Information Fig S14.* Diffusion images from 6-directions, average diffusion weighted image (DWI), b=0, color fractional anisotropy (FA) and mean diffusivity (MD) maps from the fully-sampled msEPI acquisition with MUSSELS reconstruction.

*Supporting Information Fig S15.* Diffusion images from 6-directions, root-sum-of-squares (RSoS) combination of error across the 6-directions, average DWI, color FA and MD maps from the accelerated SMS-MUSSELS reconstruction.

*Supporting Information Fig S16.* 6-direction diffusion images, RSoS error across the 6-directions, average DWI, color FA and MD maps from the magnitude-valued U-Net reconstruction.

*Supporting Information Fig S17.* Fully-sampled diffusion acquisition with MUSSELS reconstruction does not exhibit visible Nyquist ghost artifacts. Ghost-correction was performed using 1-dimensional navigators acquired for each slice and each shot individually.

*Supporting Information Fig S18.* Fully-sampled SAGE acquisition with MUSSELS reconstruction and multi-shot FLEET calibration data do not exhibit ghost artifacts. Ghost-correction was performed using 1-dimensional navigators acquired for each slice and each shot individually. 3-times scaled-up images are included to help the assessment of ghost level.